\documentstyle[11pt,IAU207_pasp,twoside,epsf]{article}
\def\edcomment#1{\iffalse\marginpar{\raggedright\sl#1\/}\else\relax\fi}
\reversemarginpar

\begin{document}

\title{Super Star Clusters in M82}
 \author{Robert W. O'Connell}
\affil{Astronomy Department, University of Virginia,  Charlottesville, VA 22903}

\begin{abstract}
M82 is the nearest starburst galaxy.  It contains two large systems
of super star clusters, one being spawned today in the active
starforming core, and one produced by an earlier starburst event which
coincided with the last orbital passage of its neighbor, M81.
The proximity of M82 makes it uniquely valuable for a wide range of studies
of massive young clusters and their environments.  
\end{abstract}

\section{Introduction:  M82, The Nearest Starburst Galaxy}

M82 is the most remarkable nearby galaxy.  Seen in the optical
continuum bands, it would strike you as a mild-mannered, if slightly
eccentric, edge-on disk system.  Its main peculiarities are a
conspicuous network of dust lanes; an unusually smooth texture at large
radius for a disk galaxy; and strange, high surface brightness knots
near its center.  However, a more careful look with narrow-band
emission line filters, first obtained by Elvius (1962) and Lynds \&
Sandage (1963), reveals a raging psychopath, with glowing plumes of
hot gas extending along the minor axis to distances of 5 kpc. 

The plumes stimulated intense scrutiny of M82 at all wavelengths from
the radio to the X-ray.  The picture which has emerged is the
following.  During the last several 100 Myr years, tidal interactions
with its large spiral companion, M81, induced a concentrated starburst
in M82 with a star formation rate of $\sim 10$ M$_\odot$ yr$^{-1}$.
Energy and gas injection from supernovae, at a rate of $\sim 0.1$
supernova yr$^{-1}$, drive a large-scale galactic wind along the minor
axis of M82.  All of the bright radio and infrared sources associated
with the active starburst (age $\la 20$ Myr) are confined within a
radius of $\sim 250$ pc of the galaxy's center.  Most of this volume
is heavily obscured by dust at optical wavelengths but is easily
probed in the infrared and radio.

Although much of the effort to date on M82 has been directed at the
starburst itself and its complex interstellar medium, M82 is also a
key fiducial for the study of super star clusters.  Not only is it the
nearest galaxy with a large system of young super star clusters (D =
3.6 Mpc, linear scale 17.5 pc/arcsec), but also it contains {\it two}
such systems: one being created in the ongoing starburst and an older
(600 Myr) one now well into its mid-life evolution.  

\section{Young Super Star Clusters in the Active Starburst}

The central regions of M82 are astonishingly complex (see Figure 1).
The first studies of the brighter knotty structures by van den Bergh (1971) and
O'Connell \& Mangano (1978, hereafter OM78) showed that these were
luminous, young, compact star clusters or cluster complexes.  The brightest
complex, 
region A, exhibited  substructure to the seeing limit and, after
a conservative extinction correction, had a remarkably bright mean
surface brightness of $\mu_{\rm V} \sim 14.5$ mag/arcsec$^2$, comparable
to those of normal galactic {\it nuclei}.  Spectral synthesis indicated it contains
stars as young as 5 Myr.  Its integrated intrinsic $M_{\rm V}$ is $ \la -17.5$. 
The M82 knots were perhaps the first recognized super star clusters,
and in fact van den Bergh (1971) used the term ``superclusters" to
describe them. 

\begin{figure}[t]   

\plotfiddle{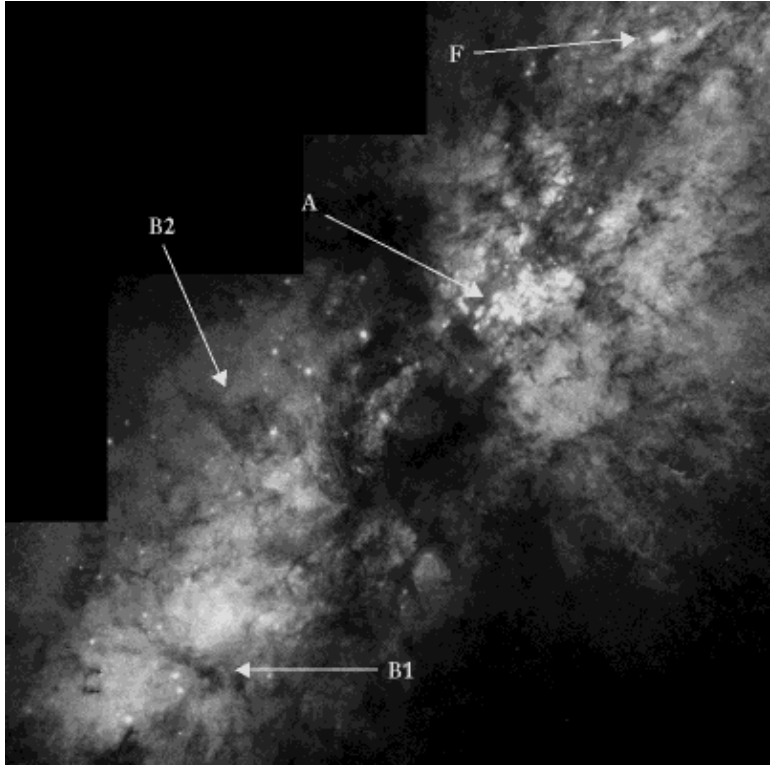}{4.0in}{0}{100}{100}{-195}{-360}

\caption{The central regions of M82 at 0.1\arcsec\ resolution.
Composite image from HST/WFPC-2 B, V, and I band frames.  About
120\arcsec\ of the major axis is shown, running diagonally across the
frame (from NE in lower left to SW in upper right).  Areas discussed
in the text are identified. A full-color version of this image is
available from STScI (PR-01-08). }

\end{figure}

HST-Planetary Camera imaging by O'Connell et al.\ (1995) resolved
complex A and its surroundings into a swarm of over 100 individual
bright clusters.  These have typical FWHM of 0.2\arcsec\ or 3.5 pc.  
The mean intrinsic brightness of the cluster sample is $M_{\rm V} \sim -11.6$
($4 \times 10 ^6$ L$_{\odot,{\rm V}}$); 
the brightest cluster has  $M_{\rm V} \sim -13.2$ (much brighter than the  most
luminous cluster in the Local Group, 30 Dor/R136, at $-11.1$).  

Infrared and radio observations, reviewed by Telesco (1988) and
Kronberg (1988), show that most of the starburst core in M82 lies
behind thick dust lanes and is not detectable in the visible bands.
Satyapal et al.\ (1997) identified 12 compact sources with
ground-based K-band imaging within the core and obtained ages in the
range 6-10 Myr for these.  Undoubtedly, HST-resolution IR imaging
would resolve these into many individual clusters, as in the case of
region A.  Recent IR spectroscopy by F\"orster Schreiber et al.\ (2001)
suggests remarkably dense packing of the ionizing sources, presumably the clusters
containing OB stars, with typical separation $\la 10$ pc.  O'Connell
et al.\ (1995) argued that region A is, in fact, part of the starburst
core which has relatively low extinction along the line of sight.
Since dust is mixed throughout the core, we can see a fractional volume
$\sim 1/\tau(\lambda)$ at any wavelength, which for the range of
estimated extinctions is $\sim$ 0.03-0.20 at V.  Over 2000 M82A-like
super star clusters could be present within the starburst core. 
There appears to be a direct link between the hot gas within region A
and the base of the H$\alpha$-bright galactic wind, implying that the
visible clusters help drive the wind.    

Controversy about the IMF within the core has persisted.  The small
M/L$_{\rm bol}$ ratio estimated from IR observations suggests that the IMF
is truncated or otherwise deficient in low mass stars (e.g. Rieke et
al.\ 1993).  However, alternative determinations of extinction and
allowance for the inhomogeneity of the core produce ratios in
agreement with normal IMF's (e.g.\ Satyapal et
al.\ 1997).  OM78 found from the emission line rotation curve that the
clusters within A have M/L$_{\rm V} \sim 0.1$ within a factor of 2,
consistent with young systems with normal IMF's.  

The most remarkable individual cluster in M82 is object F (OM78),
lying about 500 pc from region A near the western edge of the starburst
core.  With a corrected $M_{\rm V} \sim -14.5$, this was the most
luminous star cluster known until the record was broken in NGC 7252
(Schweizer \& Seitzer 1993).  Gallagher \& Smith (1999) and Smith \&
Gallagher (2001) obtained high resolution spectroscopy of F.  From
synthesis models, they were able to assign an age of $60 \pm 20$ Myr
(intermediate between the starburst and the clusters in M82 B described
below) and derived from the velocity dispersion (13.4 km/s) a mass of
$1.2 \times 10^6$ M$_{\odot}$.  The resulting M/L$_{\rm V}$ ratio is
too small by a factor of 5 to be consistent with the derived age.
After considering all the uncertainties, Smith \& Gallagher conclude
that the IMF in this cluster must be top-heavy with a lower mass limit
of 2--3 M$_{\odot}$.  Because of mass loss during stellar evolution,
the cluster is unlikely to survive more than $\sim 2$ Gyr.

\section{Intermediate Age Clusters in the Fossil Starburst}

There is now good evidence that M82 suffered at least two starburst
episodes.  Region B, extending $\sim$ 400--1000 pc NE of the galaxy's
center, has an abnormally high surface brightness and the A/F-type
absorption line spectrum associated with post-starburst systems.  OM78
proposed that this was the site of an earlier starburst in M82.  Using
HST/WFPC2 images, we (de Grijs, O'Connell, \& Gallagher 2001)
identified over 100 luminous, resolved cluster candidates in region B.
From B,V,I photometry we were able to simultaneously solve for each
cluster's age and extinction.  The clusters have intrinsic
luminosities in the range $10^{4-6}$ L$_{\odot,{\rm V}}$, significantly
fainter than the objects in region A (as expected if they are older).
The brightest cluster (visible in Fig.\ 1 near the dust lane at the
edge of region B2) has a corrected $M_{\rm V} = -10.6$.  The age
distribution in both regions B1 and B2 is broad, but in each there is
a sharp peak at age $\sim$ 600 Myr (see the de Grijs et al.\ poster at
this meeting).  Independent dynamical modeling of the HI debris in the
M81-M82-NGC3077 system by Brouillet et al.\ (1991) predicts that the
last near encounter between M81 and M82 occurred $\sim$ 500 Myr ago, coincident
with the M82 B age peak.  This suggests that cluster formation in B
was induced by the tidal passage.  Since that time, cluster formation
has been largely suppressed in region B.  

M82 is the only large system of super star clusters where the
individual members are significantly resolved by HST.  In region B we
obtain core radii in the range 2--8 pc.  The luminosity function,
corrected to a constant age of 50 Myr is similar to that in M82A and
other young super star cluster systems and consistent with
expectations for the progenitors of massive globular clusters.  It has
a power law slope of $\alpha = -1.2$, somewhat flatter than those of
most other systems, perhaps suggesting that dynamical destruction can
be detected.  

\section{Conclusion}

M82 is unique not for the magnitude of its activity but rather for its
proximity.  Starbursts of this scale are likely to be common features
of early galaxy evolution, and M82 is the nearest analogue to the
intriguing sample of star-forming galaxies recently identified (e.g.
by their Lyman dropouts) at redshifts $\ga 3$ (e.g.\ Steidel et al.
1996; Lowenthal et al.\ 1997).  M82 provides a close-up view of both
an active starburst and also, in region B, of the subsequent
post-burst phase which is apparently responsible for the ``E$+$A'' or
``quenched'' spectra which are encountered in Butcher-Oemler effect
galaxy clusters at high redshift (e.g.\ Oemler 1992; Couch et al.\
1998).  There are large numbers of super star clusters associated with
each phase.  Other nearby galaxies are known to exhibit one or another
of these features, but none affords the opportunity to study both at
such close range, or with such a wealth of correlative data,
as does M82.

Among the astrophysical issues which can be explored in M82 are the
following:  cluster structures (morphologies, gradients); the stellar
content of clusters (including CMD's in the exteriors); the IMF;
cluster dynamical evolution; propagating star formation based on
age-dating of clusters; metallicity-age-position correlations; the
structure of the ISM at high spectral resolution using clusters as
background sources; interactions between clusters in dense
environments; and cluster winds and the transition between these and
the galactic wind.


\section*{Discussion}

\noindent {\it Frogel:\,} Clarify the relative locations of (a) the
dynamical nucleus; (b) the region with all of the 600 Myr old clusters;
and (c) the origin of the H$\alpha$ gas that is being expelled from the
galaxy.\\

\noindent {\it O'Connell:\,} The dynamical center is hard to pin down
because of the complexity of the galaxy but is thought to coincide
with the 2.2$\mu$ peak, which is about 2\arcsec\ NW of region A.  The
intermediate age clusters lie in the major axis area 400-1000 pc NE of
region A.  The minor axis plume is thought to consist of gas lost from
individual massive stars and supernova remnants and a large amount of
entrained matter from the ISM.   Interestingly, a significant fraction
of the H$\alpha$ radiation from the plume is light scattered by dust grains,
apparently photons from the starburst core.  This complicates the kinematical
interpretation of the plume, and it's fair to say that a satisfactory model
has not yet emerged.  \\

\noindent {\it Whitmore:\,}  What is the size of the 600 Myr year old region?
I would have expected differential rotation to have spread it out after 600 Myr,
which is probably several orbital time scales.  Is there evidence for clusters
with similar age on the other side of the galaxy? \\

\noindent {\it O'Connell:\,} You're right that this is an interesting
problem.  The area we surveyed carefully for clusters was about 500 pc
in extent, but we examined the available HST and ground-based imagery
of the rest of the galaxy and find that bright clusters like those in
B are not common outside this region (and the core).  We don't know
how region B has managed to stay together, though this might be
consistent with a solid body rotation curve.  \\

\noindent {\it Walker:\,}  Why did the central region of M82 wait 585 Myr after
the M81/M82 interaction before lighting up? \\

\noindent {\it O'Connell:\,} Another good question, for which we like
to wave our hands and say that starbursts are probably self-quenched
after 50-100 Myr years by supernova-driven outflows, which remove any
remaining cool gas.  Of course, we see one of these now in the active core.
In the case of M82, there is a large amount of circumgalactic cold
material orbiting from the tidal encounter, and a recent infall from
this reservoir could have triggered the active event. \\

\noindent {\it Ramirez}  Could the M82 F cluster be the nucleus of a galaxy
which merged with M82?  \\

\noindent {\it O'Connell:\,} We don't think there was an actual merger.
The tidal encounter with M81 was not close.  Of course, this doesn't
rule out the possibility that M82 could have absorbed a low-mass
member of the M81 group and even could have done so relatively
recently.  Cluster F may be at the extreme, but it still seems part of
the continuum of super star cluster properties, rather than something
distinct.  \\

\noindent {\it Schweizer:\,}  The M81/M82 system is complicated, since there is
also a third galaxy (NGC 3077) participating in the interaction.  Whose model is
it that predicts an interaction with M81 about 600 Myr ago?  And did this model
specifically exclude a mass transfer, since such transfer are quite frequent
in strong interactions?  \\

\noindent {\it O'Connell:\,} The model I discussed was by Brouillet et
al.\ (1991).  It did indeed incorporate the full interaction of
the three galaxies and reproduce the HI distribution within the
M81-M82-N3077 system (though based on earlier data than the map I
showed, which was from Yun, Ho, \& Lo 1994, Nature, 372, 530).  The
model predicted a closest passage by M81 at a distance of about 21
kpc, which wouldn't necessarily induce a major gas transfer.  OM78
argued that a transfer was unlikely because the total cold material
near M82 was a large fraction (40\%) of the complement in M81 but was
about the amount expected to be tidally disrupted from M82 itself if
it had been a normal late-type disk galaxy before the encounter. 
Yun, Ho, \& Lo (1993, ApJ, 411, L17) found that the details of the HI
map near M82 likewise suggest tidal disruption rather than transfer. \\

\noindent {\it Lamers:} You found a peak in the cluster formation rate
at about 600 Myr and little evidence for cluster formation in the
intermediate age between about 15 Myr and 500 Myr.  But the cluster
disruption time in such a violent system must be small.  So clusters
formed at intermediate ages may have been disrupted, and the peak at
600 Myr was actually the last time when {\it massive} clusters were
formed that survived disruption (apart from the young burst at 15
Myr).  \\

\noindent {\it O'Connell:\,} Yes, we expected the inner regions of M82
to be a difficult environment for clusters to survive, but we were
surprised to find a significant fraction (22\%) of the clusters in
region B to be older than 1 Gyr.  By contrast, we find only 4 younger
than 100 Myr.  Although the bright background is a serious problem, we
estimated our completeness limit for cluster detection to be at
M$_{\rm V} \sim -6$, so we think we would have picked up a number of
the younger objects if they really existed, especially since they
would be brighter per unit mass.  \\

\end{document}